\documentclass[aps,prc,twocolumn,showpacs,floatfix]{revtex4-1}
\usepackage[dvipdfmx]{graphicx}
\usepackage{amssymb,amsmath,color}

\twocolumngrid
\begin{document}
\unitlength = 1mm
\title{
  Magnetoelectric behavior from cluster multipoles in square cupolas:\\ 
  Study of Sr(TiO)Cu$_4$(PO$_4$)$_4$ in comparison with Ba and Pb isostructurals
}

\author{
  Yasuyuki~Kato$^1$, Kenta~Kimura$^2$, 
  Atsushi~Miyake$^3$, Masashi~Tokunaga$^3$, Akira~Matsuo$^3$, Koichi~Kindo$^3$,
  Mitsuru~Akaki$^4$, Masayuki~Hagiwara$^4$, Shojiro~Kimura$^5$,
  Tsuyoshi~Kimura$^2$, and Yukitoshi~Motome$^1$}
\affiliation{
  $^1$Department of Applied Physics, University of Tokyo, Bunkyo, Tokyo 113-8656, Japan\\
  $^2$Department of Advanced Materials Science, University of Tokyo, Kashiwa, Chiba 277-8561, Japan\\
  $^3$Institute for Solid State Physics, The University of Tokyo, Kashiwa, Chiba 277-8581, Japan\\
  $^4$Center for Advanced High Magnetic Field Science, Graduate School of Science, Osaka University, Toyonaka, Osaka 560-0043, Japan\\
  $^5$Institute for Materials Research, Tohoku University, Sendai, Miyagi 980-8577, Japan
}

\date{\today}
%


%

\begin{abstract}
We report our combined experimental and theoretical study of magnetoelectric properties of an antiferromagnet Sr(TiO)Cu$_4$(PO$_4$)$_4$, in comparison with the isostructurals Ba(TiO)Cu$_4$(PO$_4$)$_4$ and Pb(TiO)Cu$_4$(PO$_4$)$_4$. 
The family of compounds commonly possesses a low-symmetric magnetic unit called the square cupola, which is a source of magnetoelectric responses associated with the magnetic multipoles activated under simultaneous breaking of spatial inversion and time reversal symmetries. 
Measuring the full magnetization curves and the magnetic-field profiles of dielectric constant for Sr(TiO)Cu$_4$(PO$_4$)$_4$ and comparing them with the theoretical analyses by the cluster mean-field theory, we find that the effective $S=1/2$ spin model, which was used for the previous studies for Ba(TiO)Cu$_4$(PO$_4$)$_4$ and Pb(TiO)Cu$_4$(PO$_4$)$_4$, well explains the experimental results by tuning the model parameters.
Furthermore, elaborating the phase diagram of the model, we find that the square cupolas could host a variety of magnetic multipoles, i.e., monopole, toroidal moment, and quadrupole tensor, depending on the parameters that could be modulated
by deformations of the magnetic square cupolas.
Our results not only provide a microscopic understanding of the series of the square cupola compounds, but also stimulate further exploration of the magnetoelectric behavior arising from cluster multipoles harboring in low-symmetric magnetic units.
\end{abstract}
\maketitle


\section{Introduction}
The magnetoelectric (ME) effect is a cross correlation between magnetic and electric properties of matters,
and enables us to control the electric (magnetic) polarization by the magnetic (electric) field.
The ME effect in a solid was firstly conjectured for Cr$_2$O$_3$ by Dzyaloshinskii in 1959~\cite{dzyaloshinskii1960},
and indeed observed by Astrov in 1960~\cite{astrov1960}. 
It has attracted renewed interest since the discovery of a huge ME effect in TbMnO$_3$ in 2003~\cite{kimura2003}. 
Materials hosting such a huge ME response have been extensively studied as they are potentially useful for future power-saving devices functioning without electric currents~\cite{fiebig2016}.

The necessary condition for linear ME effects (ME responses proportional to the applied magnetic and electric fields) is the absence of both spatial inversion and time reversal symmetries.  
This condition is satisfied in magnetically ordered states on noncentrosymmetric structures.
Among them particularly interesting are the systems involving noncentrosymmetric clusters made of magnetic ions, such as magnetic trimers.
In such systems, the linear ME effect is explained by magnetic multipoles defined on each cluster~\cite{gobatsevich1994,spaldin2008,kopaev2009,spaldin2013}. 
In the cluster multipole description, a spin texture on a cluster is decomposed into the magnetic monopole, toroidal moment, and quadrupole tensor, all of which are odd under 
the operations of spatial inversion and time reversal. 
Each multipole is associated with a particular ME tensor, and hence, the decomposition provides systematic understanding of the ME responses in these cluster systems.

\begin{figure}[!htb]
  \centering
  \includegraphics[width=\columnwidth]{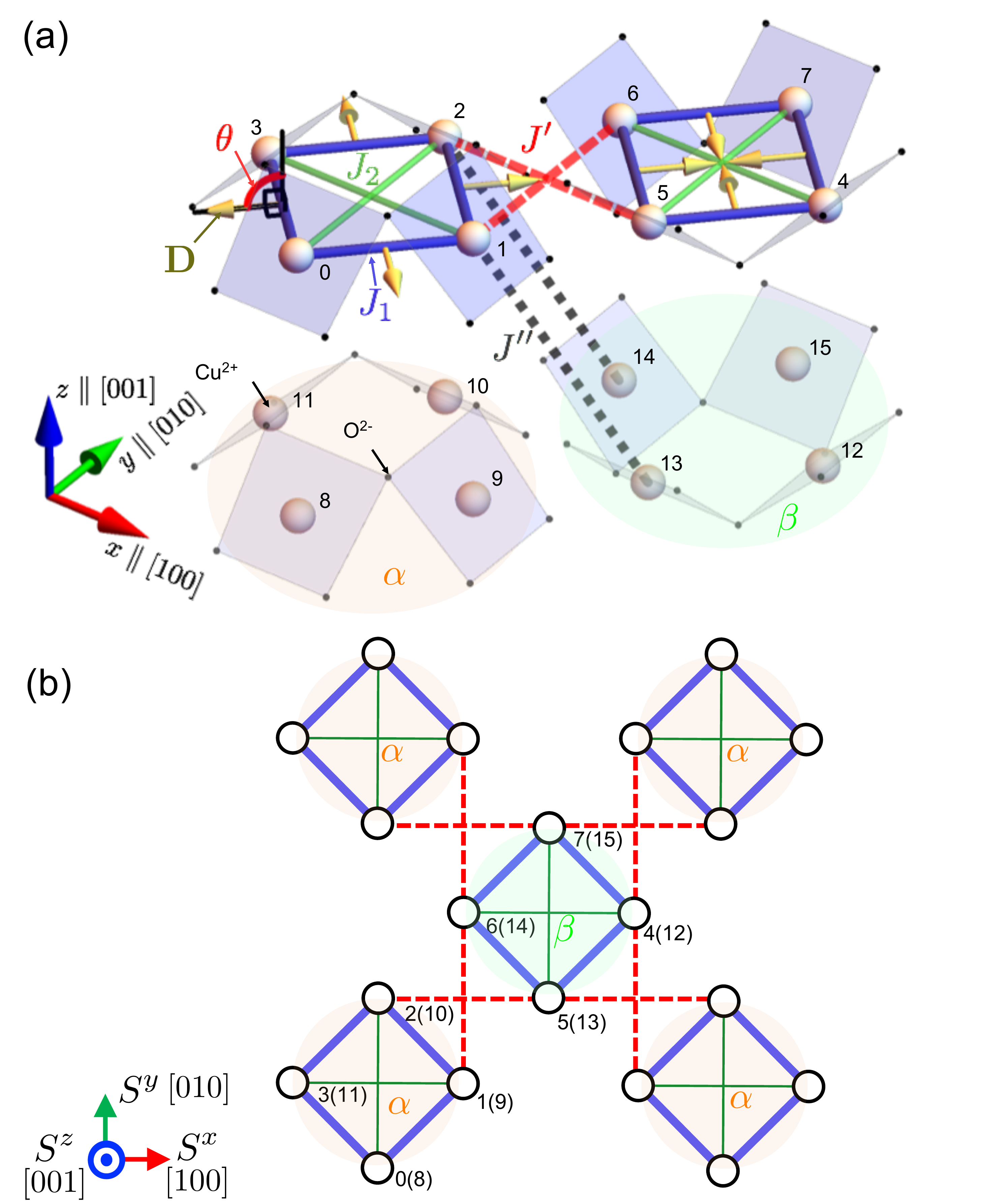}
  \caption{
    Schematic pictures of the lattice structure of $A$(TiO)Cu$_4$(PO$_4$)$_4$ ($A$TCPO, $A=$Ba, Sr, and Pb),
    which includes pairs of upward ($\alpha$) and downward ($\beta$) square cupolas composed of Cu$_4$O$_{12}$. 
    (a) Three-dimensional view
    including 16 Cu sites considered in the CMF analysis (numbered from $0$ to $15$).
    The spheres and black dots represent Cu cations and O ions, respectively.    
    The solid blue and green lines represent the intracupola couplings $J_1$ and $J_2$, respectively, while the dashed red and dotted
    gray lines are the intercupola couplings $J'$ and $J''$, respectively. 
    The yellow arrows on the $J_1$ bonds represent the DM vectors ${\bf D}_{ij}$;  
    each ${\bf D}_{ij}$ is perpendicular to the bond connecting the Cu sites $i$ and $j$ with the angle $\theta$ from the [001] axis.
    (b) Top view showing the Cu sites and the intralayer magnetic interactions.
   }
  \label{fig1}
\end{figure}

Recently, single crystals of a series of ME active insulating antiferromagnets, $A$(TiO)Cu$_4$(PO$_4$)$_4$ ($A$TCPO, $A=$ Ba, Sr, and Pb),  have been synthesized~\cite{kimura2016a,kimura2018}.
These compounds are composed of magnetic clusters Cu$_4$O$_{12}$ resembling the square cupola that is the fourth Johnson solid~\cite{johnson1966}.
Each square cupola accommodates four $S=1/2$ spin degrees of freedom from Cu$^{2+}$ cations.
The family of compounds has a quasi-two-dimensional lattice structure composed of a periodic array of the square cupolas.
More precisely, as schematically shown in Fig.~\ref{fig1}, upward ($\alpha$) and downward ($\beta$) square cupolas are alternately arranged in each layer.
In the absence of an external magnetic field, these compounds exhibit a finite-temperature ($T$) phase transition to an antiferromagnetically ordered phase where each square cupola hosts a $q_{x^2-y^2}$ quadrupole type spin texture~\cite{babkevich2017}
(the N\'eel temperature is $T_{\rm N} \simeq$ 9.5~K, 7.0~K, and  6.3~K for $A=$ Ba, Pb, and Sr, respectively).
This leads to ME responses, such as 
a dielectric anomaly at $T_{\rm N}$
in BaTCPO and SrTCPO~\cite{kimura2016,kimura2018} and  
a magnetic-field-induced net electric polarization in PbTCPO~\cite{kimura2018}. 
The difference originates from 
the way of layer stacking: the magnetic layers are stacked in a staggered manner in the Ba and Sr cases (layered antiferroic order of the $q_{x^2-y^2}$ quadrupole), 
 while in a uniform manner in the Pb case (ferroic order of the $q_{x^2-y^2}$ quadrupole).
These ME behaviors are understood in terms of the cluster multipoles of the quadrupole type.
The theoretical analyses based on a microscopic model were also reported for BaTCPO~\cite{kato2017} and PbTCPO~\cite{kimura2018b}. 
For both compounds, the cluster mean-field (CMF) theory for an effective quantum spin model 
successfully explains the ME behaviors as well as the full magnetization curves~\cite{kato2017,kimura2018b}.
More recently, the magnetic property of SrTCPO has been investigated using polycrystalline samples~\cite{islam2018}. 
However, the detailed analysis for a single crystal as well as the microscopic theory for SrTCPO has been lacked thus far.

In this paper, we investigate the ME behavior of SrTCPO by a combined experimental and theoretical analysis.
First, by experimentally measuring the magnetization curves up to full saturation for a single crystal, we identify several anomalies depending on the field direction. 
Then, we compare the experimental data with the theoretical results obtained by the CMF theory for the effective spin model, following the previous studies for BaTCPO and PbTCPO. 
We find that the theory successfully reproduces the experiment for SrTCPO as well, by tuning the model parameters.
Next, by using the same parameter set, we evaluate the dielectric constant as well as antiferromagnetic order parameters and electric polarizations, 
by which we elucidate the magnetic phase diagram at finite $T$. 
We show that the theoretical results again well agree with the experimental data of the dielectric constant measured up to 18~T. 
Thus, we conclude that our effective spin model captures the essential physics in the series of compounds for $A$=Ba, Pb, and Sr. 
In addition, we extend the theoretical analysis by interpolating the model parameters between the Sr and Ba cases and by changing the angle of the Dzyaloshinskii-Moriya (DM) vectors for the Sr parameter set. 
Although the former analysis simply connects the magnetic phases between the two compounds without any additional phases, 
the latter brings us a variety of magnetic phases, which accommodate different types of cluster multipoles: monopole, toroidal moment, and quadrupole. 
We show theoretical predictions of the ME responses in these phases, based on the cluster multipole decomposition. 
The results would stimulate further exploration of the ME effects in the family of square cupola compounds.

The structure of this paper is as follows.
In Sec.~\ref{sec:method}, we describe the experimental and theoretical methods.
The results are presented in Sec.~\ref{sec:result}.
In Sec.~\ref{sec:magcuv}, we show the experimental data of the magnetization curves, and determine the parameter set for the effective model from the comparison with the theoretical results. 
We demonstrate that the effective model well reproduces the experimental data for the ME behaviors in Sec.~\ref{sec:diel} and the finite-$T$ phase diagram in Sec.~\ref{sec:phasediagram}.
Further theoretical analyses for antiferromagnetic order parameters and electric polarizations are shown in Sec.~\ref{sec:orderparam}.
In Sec.~\ref{sec:gs}, extending the theory to a wider parameter space, we find several additional phases. 
In Sec.~\ref{sec:multipole}, we show that the distinct ME responses in these phases are explained by considering cluster multipoles.
Finally, Sec.~\ref{sec:summary} is devoted to summary and concluding remarks.
In Appendices, we show the additional theoretical results of the typical spin configurations for several phases not reported in the previous study~\cite{kato2017}
and the phase diagram for BaTCPO with the magnetic field ${\bf B}\parallel [110]$ for a complete comparison with SrTCPO.

\section{Methods}\label{sec:method}

In this section, we describe the experimental methods for the measurements of magnetization and dielectric constant. 
We also introduce the theoretical model and method for analyzing the microscopic property of the antiferromagnetic square cupola systems, $A$TCPO.

\subsection{Experimental method}
Single crystals of SrTCPO were grown by the flux method as described previously~\cite{kimura2016a}. 
Powder X-ray diffraction (XRD) measurements on crushed single crystals confirmed a single phase. 
The crystal orientation was determined by the Laue X-ray method. 
A superconducting magnet system up to 18~T and down to 1.6~K at the Tohoku University was used for measurements of dielectric properties. For dielectric measurements,  single crystals were cut into thin plates and subsequently electrodes were formed by painting silver pastes on a pair of the widest surfaces. The dielectric constant $\varepsilon$ was measured using an $LCR$ meter (Agilent E4980) at an excitation frequency of 100 kHz. Pyroelectric current was measured by an electrometer (Keithley 6517) to monitor electric polarization.
High-field magnetization in magnetic fields up to 45~T was measured at 1.4~K 
using an induction method with a multilayer pulsed magnet installed 
at the International MegaGauss Science Laboratory of Institute for Solid State Physics at The University of Tokyo. 
Multi-frequency electron spin resonance (ESR) measurements (600--1400~GHz) in pulsed magnetic fields were performed 
at the Center for Advanced High Magnetic Field Science in Osaka University
to obtain the $g$-values for the field directions along $[100]$, $[110]$ and $[001]$. 
The $g$-values were found to be isotropic within the experimental accuracy: 
$g=2.30(5)$ for all the three field directions.

\subsection{Model and theoretical method}
We consider an effective model for the $S=1/2$ spin degrees of freedom of Cu$^{2+}$ cations, 
which was first introduced for BaTCPO~\cite{kato2017} and later applied to PbTCPO~\cite{kimura2018b}. 
The model includes four dominant antiferromagnetic exchange interactions, $J_1$, $J_2$, $J'$, and $J''$, where $J_1$ and $J_2$ are intracupola exchange interactions,
and $J'$ and $J''$ are intralayer and interlayer interactions between the cupolas, respectively (Fig.~\ref{fig1}). 
In addition, we take into account the Dzyaloshinskii-Moriya (DM) interaction originating from the relativistic spin-orbit coupling on the $J_1$ bonds as well as the Zeeman coupling to an external magnetic field.
The Hamiltonian reads
\begin{eqnarray}
  \mathcal{H}
  =\sum_{\langle i,j \rangle} \left[
    J_1 {\bf S}_i \cdot {\bf S}_j
    - {\bf D}_{ij}
    \cdot {\bf S}_i \times {\bf S}_j 
    \right]
+ J_2 \sum_{\langle\langle i,j \rangle\rangle} {\bf S}_i \cdot {\bf S}_j \nonumber\\
+ J' \sum_{(i,j)} {\bf S}_i \cdot {\bf S}_j
+ J'' \sum_{((i,j))} {\bf S}_i \cdot {\bf S}_j
-g\mu_{\rm B}\sum_i {\bf B}\cdot{\bf S}_i, ~~
\label{eq:model}
\end{eqnarray}
where ${\bf S}_i=(S^x_i, S^y_i , S^z_i)$ represents the $S=1/2$ spin at site $i$, 
and the sums for $\langle i,j \rangle$, $\langle\langle i,j \rangle\rangle$,
$(i,j)$, and $((i,j))$ run over the $J_1$, $J_2$, $J'$, and $J''$ bonds, respectively.
The last term represents the Zeeman coupling with the isotropic $g$-factor $g$ 
and the Bohr magneton $\mu_{\rm B}$.
The DM interaction is characterized by the DM vector ${\bf D}_{ij}$.
For simplicity, we assume that the Cu$_4$O$_{12}$ magnetic units have the same symmetry with the perfect square cupola, $C_{4v}$.
Then, referring the Moriya rules~\cite{moriya1960}, 
we set ${\bf D}_{ij}$ in the plain perpendicular to the corresponding $J_1$ bond with a common angle $\theta_{ij}=\theta$ from the $[001]$ axis, 
and a common strength $D=|{\bf D}_{ij}|$ [the yellow arrows in Fig.~\ref{fig1}(a)]. 
Note that some features are omitted in the present model, such as the chiral twist of the square cupolas and anisotropic exchange interactions other than the DM.

In the previous analysis for BaTCPO,  
the effective model in Eq.~(\ref{eq:model}) successfully reproduces the entire magnetization curves up to above the saturation field  
and the dielectric anomaly observed at the N\'eel temperature in the low magnetic field regime with the parameter set~\cite{kato2017}: 
\begin{eqnarray}
&&J_1=1,~J_2=1/6,~
J'=1/2,~J'' = 1/100,
\nonumber\\
&& 
D = 0.7,\text{ and }\theta = 80^\circ,
 \label{eq:paramBa}
\end{eqnarray}
on the basis of an estimate of $J_1=3.03$~meV by first-principles calculations~\cite{kimura2016}.
Furthermore, by switching the sign of $J''$ from antiferromagnetic to ferromagnetic with slight changes of other parameters, 
this model is capable of reproducing the uniform manner of layer stacking with the net electric polarization appearing in PbTCPO when ${\bf B} \parallel [110]$~\cite{kimura2018b}.
In particular, the unusual sign change of the polarization observed in the high field regime is explained by the model analysis.
Through the analyses of BaTCPO~\cite{kato2017} and PbTCPO~\cite{kimura2018,kimura2018b}, 
the main origin of the ME effects is identified as the nonrelativistic exchange striction mechanism~\cite{sergienko2006}. 

In the present analysis, we optimize the model parameters to reproduce the experimental magnetization curves measured for SrTCPO, as discussed in Sec.~\ref{sec:magcuv}. 
In the calculations, following the previous analyses~\cite{kato2017,kimura2018b},
we employ the CMF method, which is suitable for cluster-based magnetic insulators. 
In the CMF method, the weak intercupola interactions ($J'$ and $J''$ terms) are dealt with 
by the conventional mean-field approximation, namely, ${\bf S}_i \cdot {\bf S}_j $ is decoupled as 
${\bf S}_i \cdot {\bf S}_j \simeq \langle {\bf S}_i \rangle \cdot {\bf S}_j + {\bf S}_i \cdot \langle {\bf S}_j \rangle - \langle {\bf S}_i \rangle \cdot \langle {\bf S}_j \rangle$, where $\langle {\bf S}_i \rangle$ is the expectation value of the spin operator ${\bf S}_i$.
On the other hand, the intracupola interactions are dealt with by the exact diagonalization, 
and therefore, quantum fluctuations in each cupola are fully taken into account.
In this paper, we consider four square cupolas allocated as shown in Fig.~\ref{fig1}(a) in the CMF method, namely we consider 16 sublattices.


\section{Results}\label{sec:result}
In this section, the results of experiments and theoretical calculations are shown.
In Sec.~\ref{sec:magcuv}, we show the experimental data of the full magnetization curves for three different directions of magnetic fields for SrTCPO, 
and determine the optimal parameter set of the theoretical model \eqref{eq:model} to reproduce the experimental results. 
We demonstrate the validity of the model for the dielectric constant and the phase diagram in Secs.~\ref{sec:diel} and \ref{sec:phasediagram}, respectively.
In Sec.~\ref{sec:orderparam}, we show the detailed analysis of the antiferromagnetic order parameters and electric polarizations in each phase. 
In Sec.~\ref{sec:gs}, we show the ground-state phase diagrams of the theoretical model 
in an extended parameter space: 
an interpolation between SrTCPO and BaTCPO, and a change of the DM angle $\theta$ for the Sr parameter set, for the latter of which we find additional phases. 
Finally, in Sec.~\ref{sec:multipole}, we investigate the ME responses for all the phases appearing in this paper
by the cluster multipole decomposition of the spin configuration of each phase. 

\subsection{Magnetization curves and model setup}\label{sec:magcuv}
\begin{figure}[!htb]
  \centering
  \includegraphics[width=\columnwidth]{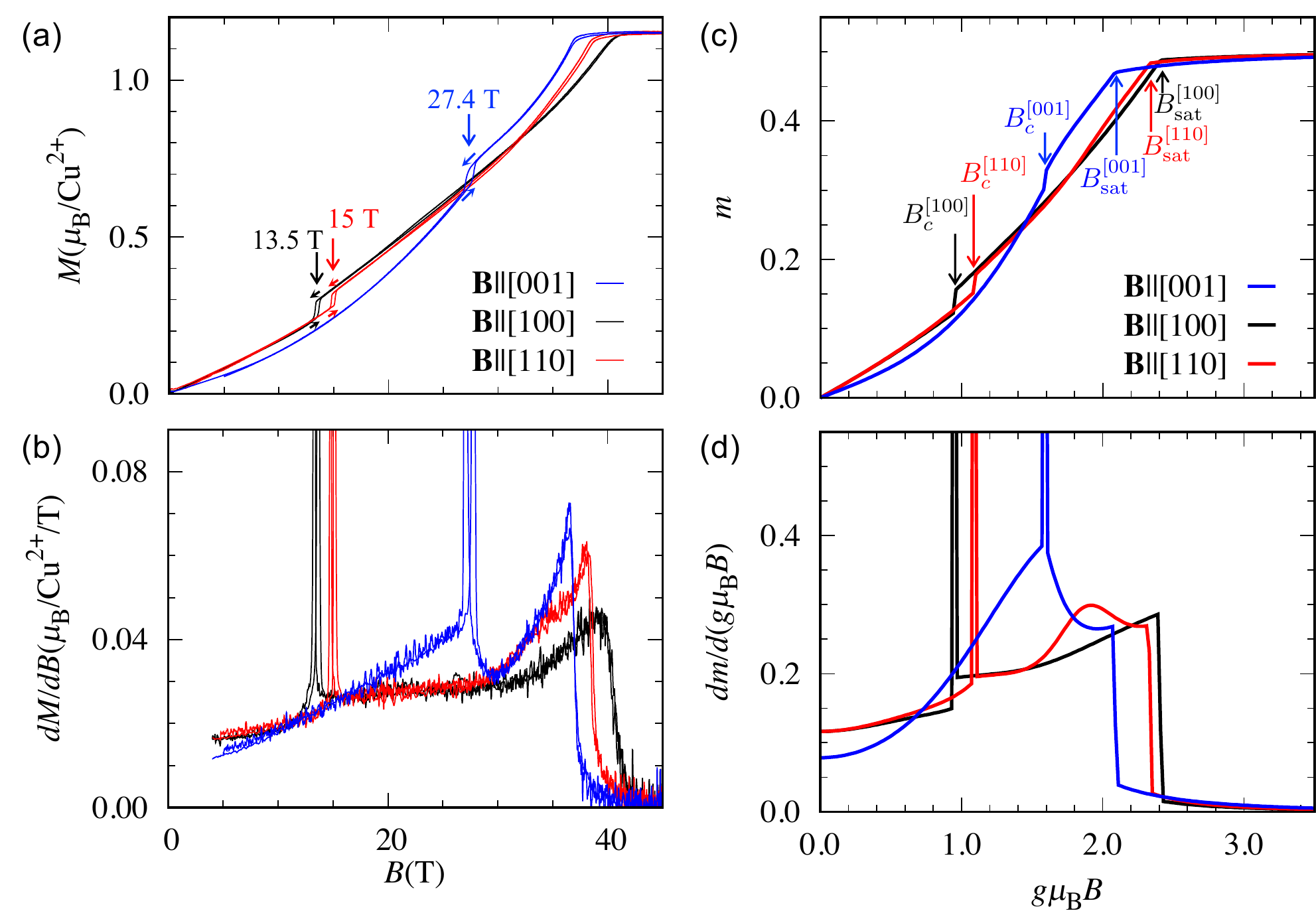}
  \caption{
  	(a,c) Magnetization curves and (b,d) their field derivatives: 
	(a,b) experimental data for SrTCPO at $T$=1.4~K and 
	(c,d) 	CMF results for the spin model~\eqref{eq:model} in the ground state with the model parameters,  
	$J_1$=0.6, $J_2$=1/6, $J'$=1/2,  $J''$=1/100, $D$=0.7, and $\theta$=$90^\circ$. 
   }
  \label{fig2}
\end{figure}

Figure~\ref{fig2}(a) shows the experimental results of full magnetization curves at $T=1.4$~K $< T_{\rm N}$ for the magnetic field 
applied along the $[001]$, $[100]$, and $[110]$ directions. 
In the low field region ($B=|{\bf B}| \lesssim 5$~T), the slope of the magnetization $M$ is smaller for the out-of-plane field (${\bf B} \parallel [001]$) 
than for in-plane fields (${\bf B} \parallel [100]$ and $[110]$), similarly to BaTCPO~\cite{kato2017} or PbTCPO~\cite{kimura2018b}. 
In the higher field region, we find a jump-like anomaly in $M$ with a small hysteresis for all the ${\bf B}$ directions. 
The critical fields, defined as a central value of $B$ for each hysteresis, are $B^{[100]}_c \simeq 13.5$~T, $B^{[110]}_c \simeq 15.0$~T, and 
$B^{[001]}_c \simeq 27.4$~T for ${\bf B}\parallel[100]$, $[110]$, and $[001]$, respectively.
These anomalies are more clearly seen in the field derivative $dM/dB$ in Fig.~\ref{fig2}(b). 
Above $B \simeq 40$~T, 
the magnetization for all the ${\bf B}$ directions shows a saturation at $\sim$1.15~$\mu_{\rm B}/{\rm Cu}^{2+}$. 
The saturation-magnetization values are corrected by the $g$-values determined by the ESR.
We note that $dM/dB$ shows a hump at $B \simeq 35$~T only for ${\bf B}\parallel[110]$ as shown in Fig.~\ref{fig2}(b).

A significant difference between the magnetization curves of SrTCPO 
and those of BaTCPO and PbTCPO is found in the relative magnitude of $B^{[001]}_c$ and $B^{[100]}_c$, namely,
 $B^{[001]}_c > B^{[100]}_c$ for SrTCPO while
 $B^{[001]}_c < B^{[100]}_c$ for BaTCPO and PbTCPO.
Furthermore, the ratio of the critical field to the saturation field, $b^{[001]}_c \equiv B^{[001]}_c / B^{[001]}_{\rm sat}$, is much larger:
$b^{[001]}_c \sim 0.75$ for SrTCPO 
while $b^{[001]}_c \sim 0.2$ for BaTCPO~\cite{kato2017} and $b^{[001]}_c \sim 0.3$ for PbTCPO~\cite{kimura2018b}.
We find that these aspects are reproduced simply by taking a smaller $J_1$ as $J_1 \lesssim 0.6$, while  
keeping the other parameters as those for BaTCPO in Eq.~\eqref{eq:paramBa}. 
We note that the smaller $J_1$ is also reasonable to reproduce the smaller saturation fields $\sim 40$~T compared to $\sim 60$~T in BaTCPO~\cite{kato2017}. 
At the same time, however, we find that the parameter change leads to an additional phase transition not observed in experiments 
in the higher field regime for ${\bf B}\parallel [110]$.
This is remedied by a slight increase of $\theta$. 
Consequently, 
we obtain the optimal parameter set for SrTCPO by adjusting only $J_1$ and $\theta$ as
\begin{eqnarray}
&&J_1=0.6
~\text{ and }\theta = 90^\circ, 
\label{eq:paramSr}
\end{eqnarray}
from Eq.~\eqref{eq:paramBa} for BaTCPO.

The main difference of the model parameters between SrTCPO and BaTCPO is in the magnitude of the nearest-neighbor exchange interaction $J_1$;
$J_1$ for SrTCPO is taken as 60\% of that for BaTCPO.
The parameter change is consistent with the fact that both the saturation field and the Curie Weiss temperature of SrTCPO are approximately 2/3 of those of BaTCPO in experiments~\cite{kimura2018}.
We note that $J_1$ was estimated to be $\sim 3$~meV commonly for the Sr and Ba cases in the first-principles calculations~\cite{kimura2016,kimura2018}, but the values cannot explain the experimental observations within the model analysis.

In Figs.~\ref{fig2}(c) and \ref{fig2}(d), we show the theoretical results for the magnetization curves 
and their field derivatives at zero $T$, respectively.
The entire magnetization curves are well reproduced by the optimal parameter set, in the following aspects:
(i) $B^{[001]}_{\rm sat}<B^{[110]}_{\rm sat}<B^{[100]}_{\rm sat}$, 
(ii) $B^{[100]}_{c} <B^{[110]}_{c} <B^{[001]}_{c}$, 
(iii) $b^{[001]}_c \sim 0.75$,  
(iv) the field derivative for out-of-plane field (${\bf B}\parallel [001]$) lower than that for in-plane field (${\bf B}\parallel [100]$ or $[110]$) in the low field regime,
and
(v) a hump near the saturation in the field derivative for ${\bf B}\parallel [110]$.


\subsection{Dielectric anomaly}\label{sec:diel}
\begin{figure}[!htb]
  \centering
  \includegraphics[ width=\columnwidth,trim=10 0 5 0,clip]{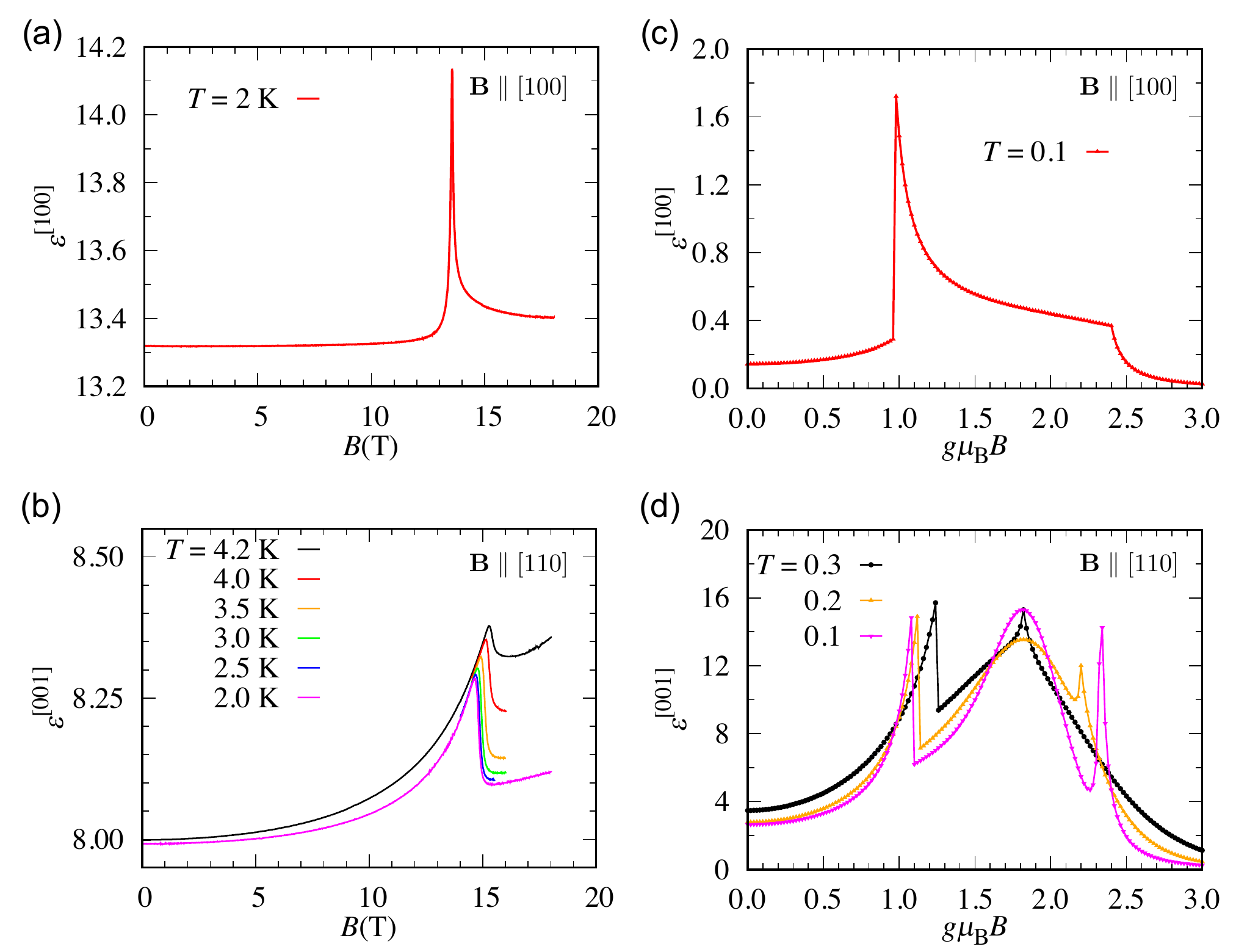}
  \caption{
  Magnetic field dependence of the dielectric constant at low $T$ 
  obtained in (a,b) experiments for SrTCPO and 
  (c,d) theoretical calculations for the model~\eqref{eq:model}. 
  The magnetic and electric fields are taken as 
  (a,c) ${\bf B}\parallel {\bf  E}\parallel [100]$ and 
  (b,d) ${\bf B}\parallel [110]$ and ${\bf E}\parallel [001]$. 
  The parameter set for SrTCPO (see the caption of Fig.~\ref{fig2}) and $\Delta E = 0.0025$ [see Eq.~\eqref{eq:epsilon}] are used in (c,d).
   }
  \label{figure3}
\end{figure}
\begin{figure}[!htb]
  \centering
  \includegraphics[width=\columnwidth]{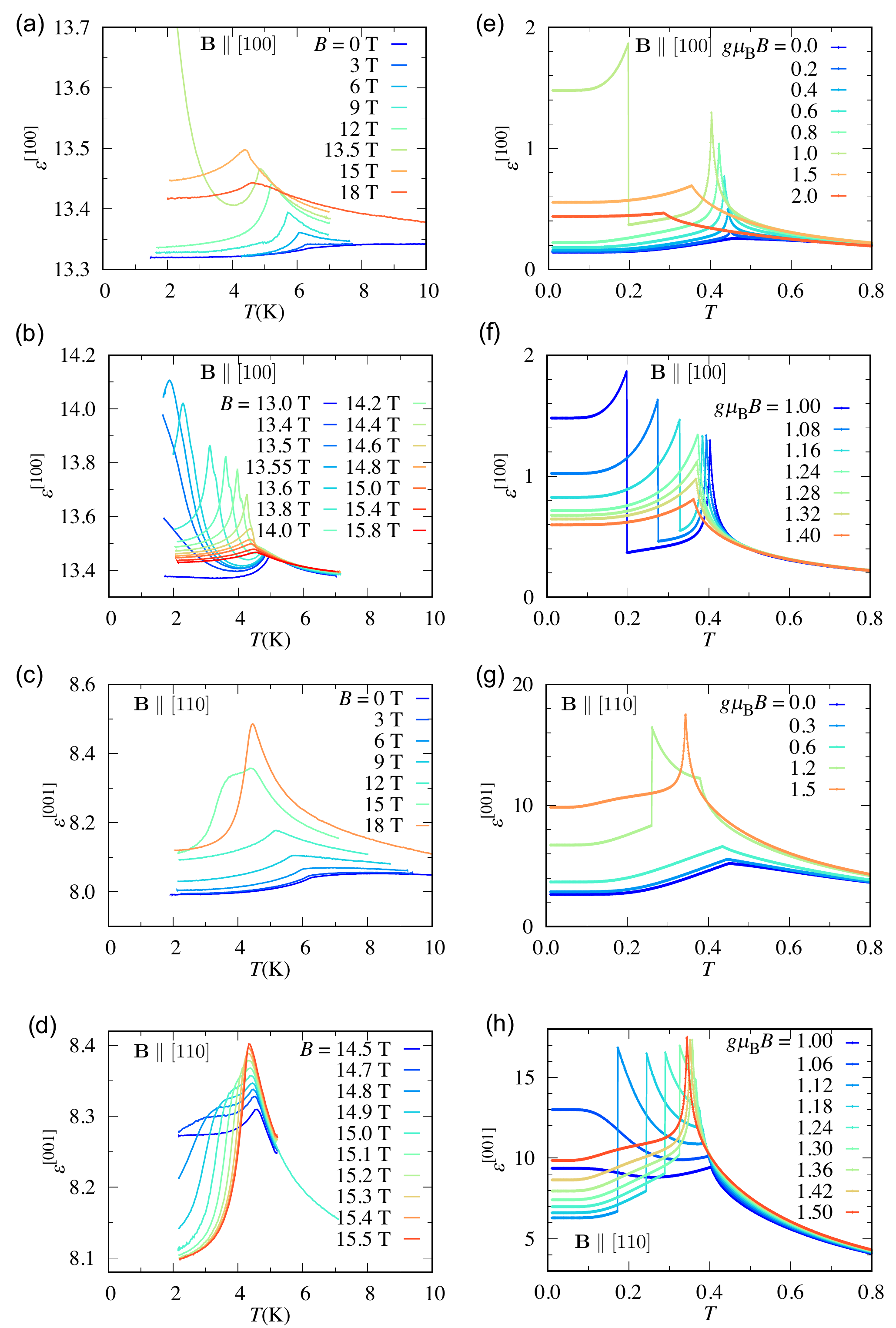}
  \caption{
  $T$ dependence of the dielectric constant obtained in (a--d) experiments for SrTCPO and (e--h) theoretical calculations for the model~\eqref{eq:model}, for various magnetic field strength: 
  (a,b,e,f) ${\bf B}\parallel {\bf E}\parallel [100]$ and (c,d,g,h) ${\bf B}\parallel [110]$ and ${\bf E}\parallel [001]$. 
  (b,d,f,h) are the results close to the critical fields. 
  The same parameters as in Fig.~\ref{figure3} are used in the theory. 
   }
  \label{figure4}
\end{figure}

Since the maximum field of 18~T available in the present dielectric measurements cannot access the critical field for ${\bf B}\parallel [001]$, we performed the dielectric measurements only in ${\bf B}\parallel [100]$ and ${\bf B}\parallel [110]$.
Figures~\ref{figure3}(a) and \ref{figure3}(b) show the experimental data of the dielectric constant at low $T$
measured for SrTCPO in ${\bf B}\parallel {\bf E}\parallel [100]$, and ${\bf B}\parallel [110]$ and ${\bf E}\parallel[001]$ up to $B=
18$~T, respectively 
(${\bf E}$ is the electric field).
The dielectric constant exhibits sharp anomalies at the magnetic fields where the magnetization changes discontinuously.
Note that the pyroelectric current measurement does not detect any signal indicative of an onset of a macroscopic electric polarization associated with these dielectric anomalies.

For comparison, in Figs.~\ref{figure3}(c) and \ref{figure3}(d), we show the corresponding theoretical results computed by the CMF method with the parameter set for SrTCPO (see the caption of Fig.~\ref{fig2}).
Note that the experimental data are limited to the field range below 18~T, which roughly corresponds to $g\mu_{\rm B}B<1.5$ in the theoretical results.
In the CMF method, 
introducing an electric field term to the Hamiltonian as $\mathcal{H}-{\bf E} \cdot {\bf P}$,
the dielectric constant is evaluated as
\begin{equation}
\varepsilon^{[abc]} =
\frac{
\langle {\bf P}\cdot {\bf n} \rangle_{{\bf E}=\Delta E {\bf n}}
- \langle {\bf P}\cdot {\bf n} \rangle_{{\bf E}= 0}
 }{
 \Delta E
 },
 \label{eq:epsilon}
\end{equation}
with a sufficiently small $\Delta E$ where ${\bf n}$ is the normalized vector directing $[abc]$.
Following the previous studies~\cite{kato2017,kimura2018b},
we consider the electric polarization induced by the exchange striction mechanism~\cite{sergienko2006}: 
the net electric polarization is defined as
\begin{equation}
 {\bf P} = \sum_{\langle i,j\rangle}  {\bf n}_{ij} \langle {\bf S}_i \cdot{\bf S}_j \rangle,
\label{eq:P}
\end{equation}
where ${\bf n}_{ij}$ is the normalized vector from the center of the $ij$ bond to an O site shared by the CuO$_4$ squares for the Cu sites $i$ and $j$~\cite{kato2017}.
We note that $\varepsilon^{[abc]}$ in Eq.~\eqref{eq:epsilon} represents not the entire but major contribution to the dielectric constant from the spin texture through the exchange striction mechanism in an arbitrary unit.
The theoretical curves for $g\mu_{\rm B}B\lesssim 1.5$ qualitatively well reproduce the experimental results, not only the sharp anomalies but also
the asymmetric shapes of the peaks. 
In addition, they also reproduce further details of the data: the decrease (increase) while increasing $B$ after the peaks for ${\bf B}\parallel [100]$ ($[110]$), 
and the reduction and shift of the peak while increasing $T$ for ${\bf B}\parallel [110]$. 

Figures~\ref{figure4}(a-d) show the experimental data for the $T$ dependence of the dielectric constant for different magnetic fields, 
and Figs.~\ref{figure4}(e-h) are the corresponding theoretical results. 
Again, the theoretical curves well reproduce the experimental results; 
for instance, in comparison of Figs.~\ref{figure4}(a,b) and \ref{figure4}(e,f) for ${\bf B}\parallel{\bf E}\parallel [100]$, 
the increase of the dielectric anomaly while increasing $B$ toward the critical field $B^{[100]}_c \simeq 13.5$~T in the low field regime, and the nondivergent cusplike feature for higher fields.
In the same way, Figs.~\ref{figure4}(c,d) and \ref{figure4}(g,h) for ${\bf B}\parallel [110]$ and ${\bf E}\parallel [001]$ show good correspondence between the experimental measurements and the theoretical calculations;
the increase of the dielectric anomaly in the low field regime,
and the sharper anomaly for higher fields.


\subsection{Finite-temperature phase diagram}\label{sec:phasediagram}

\begin{figure}[!htb]
  \centering
  \includegraphics[width=\columnwidth,trim = 8 0 7 0,clip]{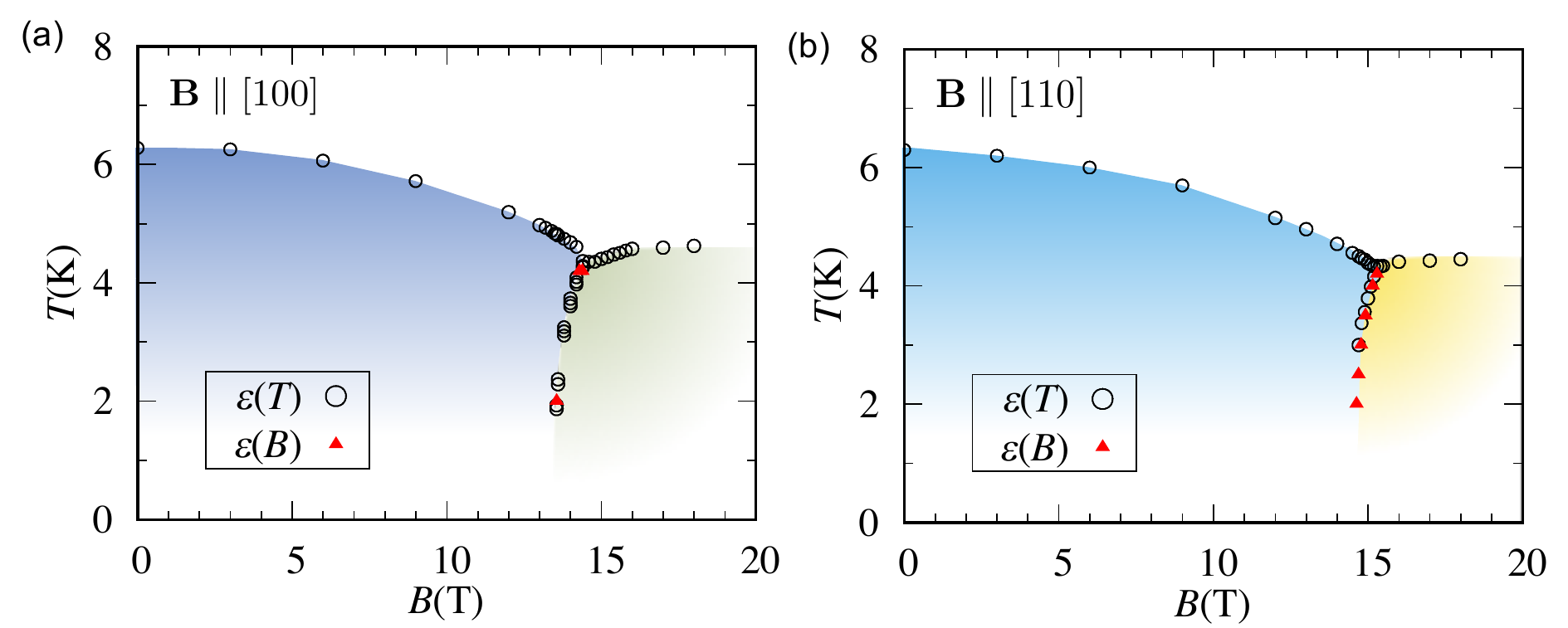}
  \caption{
	Finite-$T$ phase diagrams obtained in experiment for the magnetic field (a) ${\bf B}\parallel [100]$ 
	and (b) ${\bf B}\parallel [110]$. 
	The phase boundaries $\varepsilon(T)$ and $\varepsilon(B)$ are determined by the peak positions of the dielectric constant while changing $T$ and $B$ , respectively.
   }
  \label{figure5}
\end{figure}
\begin{figure*}[!htb]
  \centering
  \includegraphics[width=\textwidth]{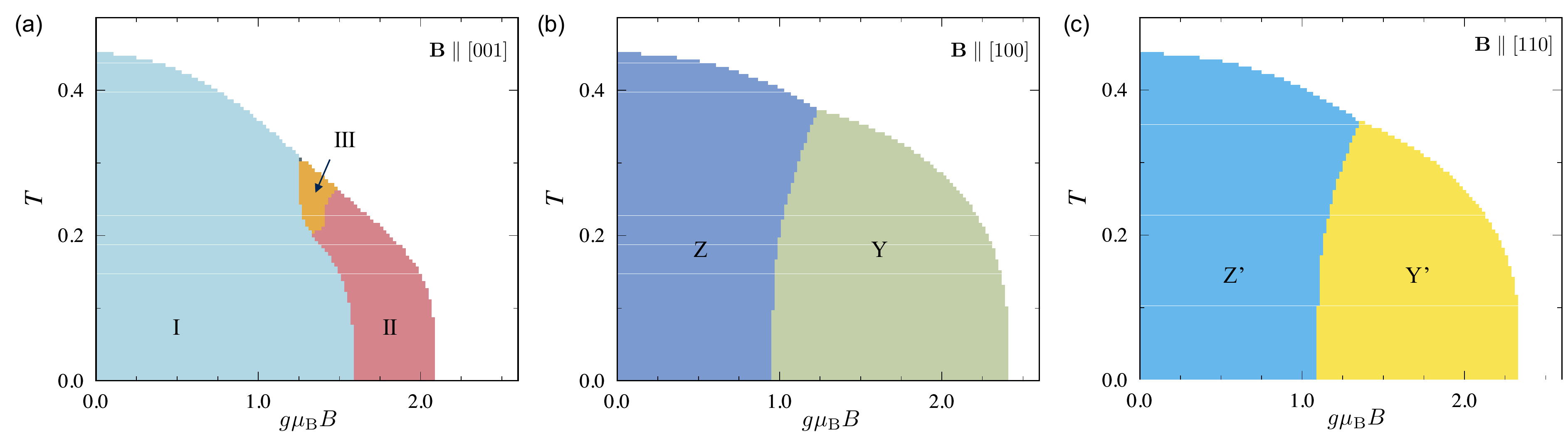}
  \caption{
  Finite-$T$ phase diagrams for the magnetic field
  (a) ${\bf B}\parallel [001]$, (b) ${\bf B}\parallel [100]$, 
  and (c) ${\bf B}\parallel [110]$ computed by the CMF method with the parameter set for SrTCPO (see the caption of Fig.~\ref{fig2}).
   }
  \label{figure6}
\end{figure*}

We summarize the experimental phase diagrams in Fig.~\ref{figure5} by plotting the peak positions of the dielectric constant. 
For both cases with ${\bf B}\parallel [100]$ and ${\bf B}\parallel [110]$, the critical temperatures separating the high-$T$
paramagnet and the low-$T$
ordered phase are reduced by increasing the magnetic field in the low field region. 
The system exhibits a phase transition at $B_c^{[100]} \simeq 13.5$~T and $B_c^{[110]}\simeq 15$~T at low $T$, and the critical fields slightly increase while raising $T$. 
The critical temperatures of the high field phase to the paramagnetic state show a small increase in the narrow field region of the measurement.

We show the finite-$T$ phase diagram obtained by the CMF method in Fig.~\ref{figure6}. 
Figures~\ref{figure6}(b) and \ref{figure6}(c) correspond to the experimental results in Fig.~\ref{figure5}. 
We find that the phase diagrams for ${\bf B}\parallel [100]$ and ${\bf B}\parallel [110]$ are similar to each other; 
we call the low-field ordered phase Z (Z') and the high field one Y (Y') for ${\bf B}\parallel [100]$ ($[110]$). 
The results indicate that our theory well reproduces the experimental results in Fig.~\ref{figure5}, except for the small enhancement of the critical temperature in the high field phase. 
This discrepancy might be reconciled by taking into account the fluctuation effect beyond the CMF approximation which may play an important role in the phase competing region. 
Based on the good agreement between the experiment and theory, 
we identify the low field phases in experiments as Z and Z' and the high field phases as Y and Y'. 
We will discuss the order parameters and electric polarizations in these phases in Sec.~\ref{sec:orderparam}. 

In addition, we also show the phase diagram for ${\bf B}\parallel [001]$ in Fig.~\ref{figure6}(a), 
in which the high-field phases (II and III) are
not accessible in the present dielectric experiments.
The phase diagram is similar to that for BaTCPO~\cite{kato2017}: 
the stabilized phases are common, including the hidden phase III. 
We note that the phase diagrams in Figs.~\ref{figure6}(b) and \ref{figure6}(c) are also similar to those for BaTCPO 
(see Ref.~\cite{kato2017} for ${\bf B}\parallel [100]$ and Appendix~\ref{app:PD:BaTCPO} for ${\bf B}\parallel [110]$).

\subsection{Order parameters and electric polarizations}\label{sec:orderparam}
\begin{figure}[!htb]
  \centering
  \includegraphics[width=\columnwidth, trim=10 0 0 0, clip]{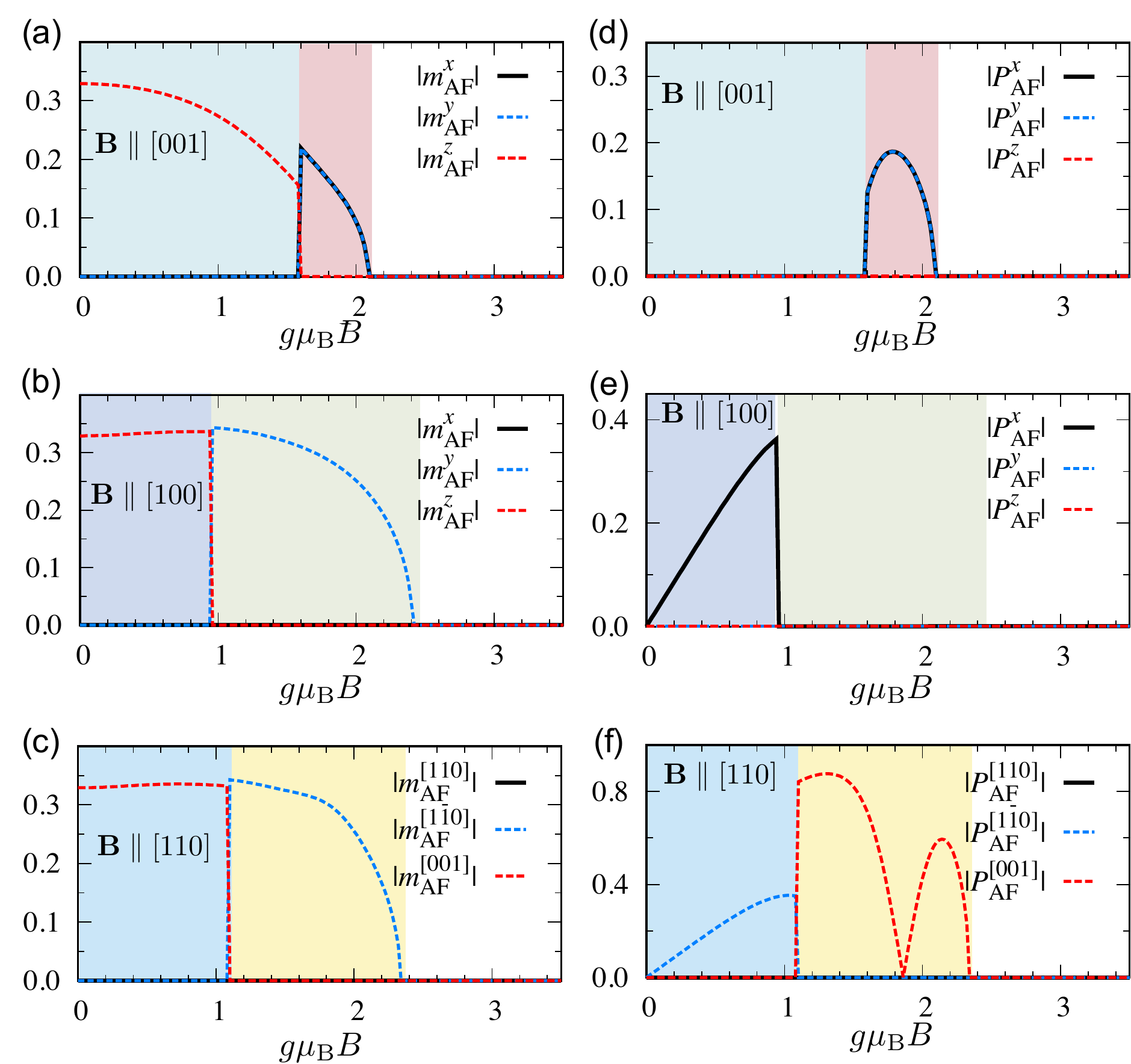}
  \caption{
  $B$ dependence of (a-c) the antiferromagnetic order parameter ${\bf m}_{\rm AF}$ [Eq.~(\ref{eq:m_AF})] and
  (d-f) the interlayer-staggered component of the electric polarization, ${\bf P}_{\rm AF}$ [Eq.~(\ref{eq:P_AF})], 
  for  (a,d) ${\bf B}\parallel [001]$,
         (b,e) ${\bf B}\parallel [100]$,
   and (c,f) ${\bf B}\parallel [110]$.
   }
  \label{figure7}
\end{figure}

Based on the similarity of the phase diagrams, here 
we analyze the theoretical results for SrTCPO by the antiferromagnetic order parameters used in the study of BaTCPO~\cite{kato2017}: 
\begin{equation}
{\bf m}_{\rm AF} \equiv \frac{1}{N_{\rm spin}}\sum_\ell (-1)^\ell p_\ell \langle {\bf S}_\ell \rangle,
\label{eq:m_AF}
\end{equation}
where $p_\ell = +1(-1)$ for the upper (lower) layer in Fig.~\ref{fig1}(a), and $N_{\rm spin}$ is the number of spins.
Figures~\ref{figure7}(a-c) show the magnetic field dependence of ${\bf m}_{\rm AF}$ at zero $T$ for the three different field directions~\footnote{We found a factor 2 missing in the plot of ${\bf m}_{\rm AF}$ in Ref.~\cite{kato2017}.}.
While only the $z$ component of the order parameter is nonzero 
($m^z_{\rm AF} \neq 0$  and $m^{x,y}_{\rm AF} = 0$) for the low field phase including $B=0$,
the orientation of ${\bf m}_{\rm AF}$ changes 
to the perpendicular direction to the $z$ axis 
through a first-order phase transition with the magnetization jump: 
$|m_{\rm AF}^x|=|m_{\rm AF}^y|\neq 0$ for ${\bf B}\parallel [001]$, 
$m_{\rm AF}^y \neq 0$ for ${\bf B}\parallel [100]$, and 
$m_{\rm AF}^{[1\bar{1}0]} \neq 0$ for ${\bf B}\parallel [110]$.
${\bf m}_{\rm AF}$ vanishes continuously at the saturation field for all the directions.

Figures~\ref{figure7}(d-f) show the field dependence of the staggered component of the electric polarization,
${\bf P}_{\rm AF}$, computed based on the exchange striction mechanism~\cite{sergienko2006}.  
In the present system, a ferroelectric polarization can appear in each layer, but the direction is antiparallel between the neighboring layers, resulting in the vanishing net polarization. 
Thus, we define the interlayer-staggered component as~\footnote{
We found a mistake in the definition of ${\bf P}_{\rm AF}$ in Ref.~\cite{kato2017}: $q_i$ should be omitted.}
\begin{equation}
{\bf P}_{\rm AF} = \sum_{\langle i,j\rangle} p_i {\bf n}_{ij} \langle {\bf S}_i \cdot{\bf S}_j \rangle.
\label{eq:P_AF}
\end{equation}
${\bf P}_{\rm AF}$ behaves differently for three field directions: 
$|P_{\rm AF}^x|=|P_{\rm AF}^y| \neq 0$ in the phase II for ${\bf B}\parallel [001]$, 
$P_{\rm AF}^x \neq 0$ in Z for ${\bf B}\parallel [100]$, and 
$P_{\rm AF}^{[1\bar{1}0]} \neq 0$ in Z' and $P_{\rm AF}^{[001]} \neq 0$ in Y' for ${\bf B}\parallel [110]$. 
Note that $P_{\rm AF}^{[001]}$ changes its sign in the phase Y'. 
Similar behavior was found in PbTCPO as a sign change of the net electric polarization parallel to [001]~\cite{kimura2018b}.

The results for ${\bf m}_{\rm AF}$ and ${\bf P}_{\rm AF}$ are summarized in Table~\ref{tab1}. 
The table includes other phases found in Sec.~\ref{sec:gs} by changing the model parameters.

\begin{table}[!htb]
\centering
\begin{tabular}{cccccc}
  \hline
  \hline
  &   & 
  ${\bf m}_{\rm AF}$  & 
  ${\bf P}_{\rm AF}$  & 
  ${\bf P}$  &  notes
  \\
  \hline	
	&I  		& $[001]$			& -				& -		&  \\
	&II  		& $[110]/[1\bar{1}0]$	& $[1\bar{1}0]/[110]$	& -		& ${\bf m}_{\rm AF} \perp {\bf P}_{\rm AF}$ \\
${\bf B}\parallel [001]$
	&III  		& $[100]/[010]$		& $[100]/[010]$		& -		& ${\bf m}_{\rm AF} \parallel {\bf P}_{\rm AF}$ \\
	&IV  		& -				& -				& $[001]$	& ${\bf P} \parallel {\bf B}$ \\
	&V  		& -              		& -				& -		& \\
 \hline
	&Z  		& $[001]$			& $[100]$			& -		& ${\bf m}_{\rm AF} \perp {\bf P}_{\rm AF}$ \\
	&Y  		& $[010]$			& -				& -		&  \\
${\bf B}\parallel [100]$
	&M  		& -                          	& -				& $[100]$	& ${\bf P} \parallel {\bf B}$  \\
	&T  		& -    			& -		 		& $[ab0]$	& $P^x \neq P^y$  \\
	&S  		& $[001]$			& $[ab0]$	 		& -	& $P^x_{\rm AF} \neq P^y_{\rm AF}$  \\
 \hline
	&Z'  		& $[001]$			& $[1\bar{1}0]$		& -		& ${\bf m}_{\rm AF} \perp {\bf P}_{\rm AF}$ \\
${\bf B}\parallel [110]$
	&Y'  		& $[1\bar{1}0]$		& $[001]$			& -		& ${\bf m}_{\rm AF} \perp {\bf P}_{\rm AF}$ \\
	&M'  		& -				& -				& $[110]$	& ${\bf P} \parallel {\bf B}$ \\
  \hline  
  	$B=0$ &($\theta > \theta_c$)	& $[001]$ 		& - 			& - 	&   \\
  	$B=0$ &($\theta < \theta_c$)	& -	 		& - 			& - 	&   \\
	  \hline    \hline
\end{tabular}
\hspace{5mm} 
\caption{
 Direction of the antiferromagnetic order parameter ${\bf m}_{\rm AF}$ [Eq.~\eqref{eq:m_AF}], 
 the interlayer-staggered component of the electric polarization, ${\bf P}_{\rm AF}$ [Eq.~\eqref{eq:P_AF}],
 and the net electric polarization ${\bf P}$ [Eq.~\eqref{eq:P}] in each phase.
 The symbol ``-'' indicates that the order parameter vanishes. 
 $\theta_c$ is the critical angle at $B=0$: $\theta_c = 12.5\pm 0.05 ^\circ$ (see the text for details). 
  \label{tab1}
}
\end{table}

\subsection{Ground-state phase diagram in an extended parameter space}\label{sec:gs}

Thus far, we have discussed the model in Eq.~\eqref{eq:model} with the parameter set for SrTCPO. 
In this section, we extend the parameter space and try to find other interesting ME behaviors for future material investigation.

\begin{figure}[!htb]
  \centering
  \includegraphics[width=\columnwidth, trim=10 0 0 0, clip]{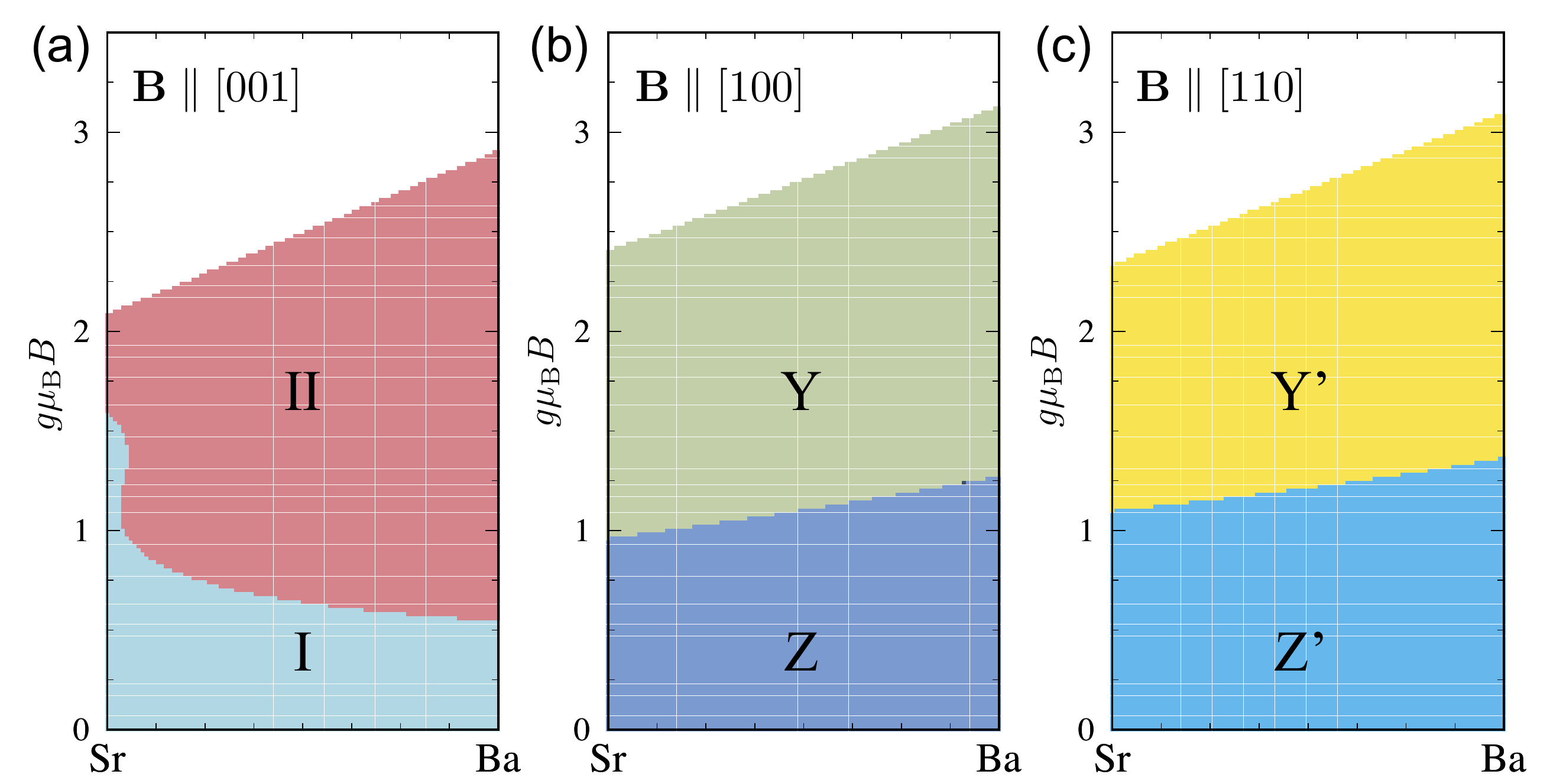}
  \caption{
	Ground-state ($T=0$) phase diagrams for the parameter sets linearly interpolated between 
	$(J_1,\theta)=(0.6, 90^\circ)$ for Sr(TiO)Cu$_4$(PO$_4$)$_4$ and
	$(J_1,\theta)=(1, 80^\circ)$ for Ba(TiO)Cu$_4$(PO$_4$)$_4$. 
	The other model parameters are fixed at $J_2 = 1/6$, $J'=1/2$, $J''=1/100$, and $D=0.7$.
	The magnetic field direction is
        (a) ${\bf B} \parallel [001]$, (b) ${\bf B} \parallel [100]$, and (c) ${\bf B} \parallel [110]$.
   }
  \label{figure8}
\end{figure}

First, considering a solid solution of the Sr and Ba compounds, we study the interpolation between the parameter sets for SrTCPO and BaTCPO. 
Figure~\ref{figure8} shows the ground-state phase diagrams computed by changing the parameters continuously between SrTCPO and BaTCPO. 
For simplicity, here we interpolate $J_1$ and $\theta$ linearly between $(J_1,\theta) = (0.6, 90^\circ)$ for SrTCPO and $(J_1,\theta) = (1, 80^\circ)$ for BaTCPO. 
For all the three field directions, the phase diagrams change continuously, without any additional phases.
The experimentally-observed jumps in the magnetization curves of SrTCPO in Fig.~\ref{fig1} are identified as the ME transitions between I and II for ${\bf B} \parallel [001]$,
between Z and Y for ${\bf B}\parallel [100]$, and between Z' and Y' for ${\bf B}\parallel [110]$ as those of BaTCPO.
(See Appendix~\ref{app:PD:BaTCPO} for the phase diagrams for BaTCPO with ${\bf B}\parallel [110]$.
The phase diagrams for the other field directions are in Ref.~\cite{kato2017}.)
Thus, although the phase boundary between I and II in Fig.~\ref{figure8}(a) shows a rapid and reentrant change for slight doping of Ba, 
our results imply no qualitatively new ME phase for a solid solution (Sr,Ba)TCPO.

\begin{figure*}[!htb]
  \centering
  \includegraphics[width=\textwidth]{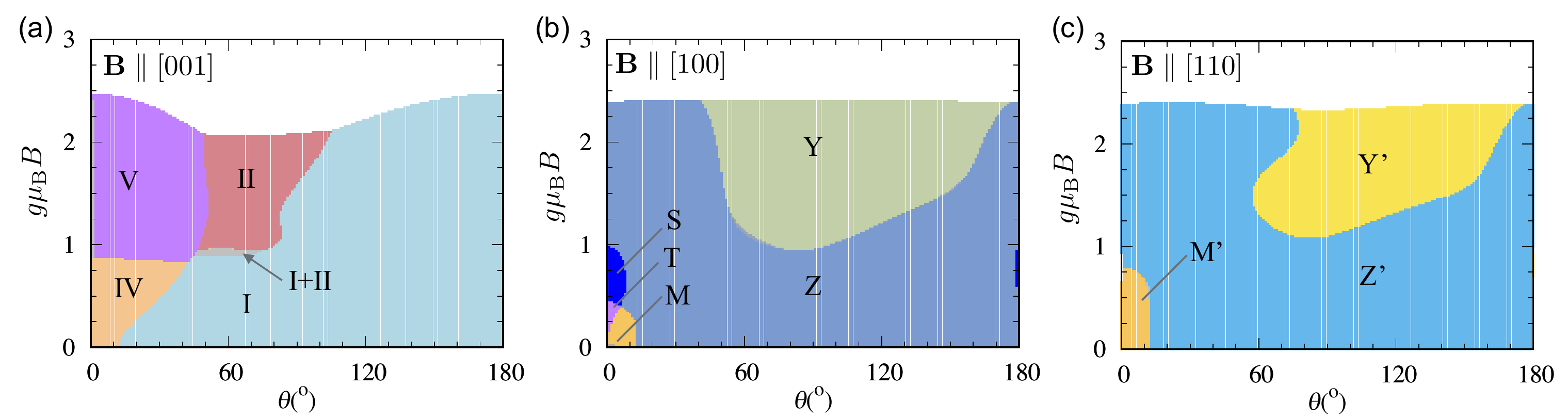}
 \caption{
  Ground-state ($T=0$) phase diagrams 
  by changing the DM angle $\theta$ with the parameter set for SrTCPO in 
  (a) ${\bf B}\parallel [001]$,  (b) ${\bf B}\parallel [100]$, and (c) ${\bf B}\parallel [110]$.
  The colored regions represent different magnetically ordered phases, while the white regions are the forced ferromagnetic phases. 
  The gray region I+II in (a) indicates a mixed phase. There are also narrow mixed regions near other phase boundaries although they are not seen clearly in the figures.
   }
  \label{figure9}
\end{figure*}
\begin{figure}[!htb]
  \centering
  \includegraphics[width= \columnwidth,trim = 30 0 0 0,clip]{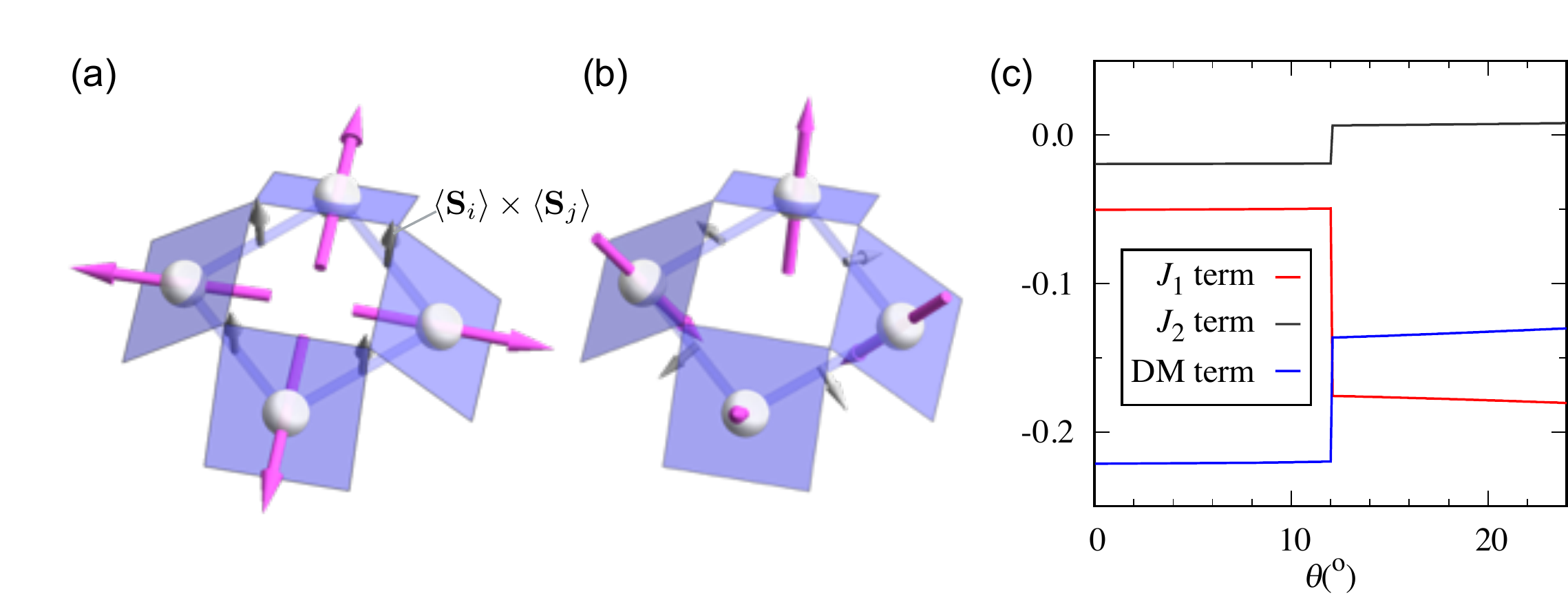}
  \caption{
  Spin configurations in a square cupola of (a) flux type at $\theta =8^{\circ} <\theta_c$ and (b) $q_{x^2-y^2}$ quadrupole type at $\theta=90^{\circ} >\theta_c$ for $B=0$. 
  The magenta arrows denote the spins and the gray arrows on the bonds represent $\langle {\bf S}_i \rangle \times \langle {\bf S}_j \rangle$.
  (c) Contributions to the energy density from the $J_1$, $J_2$, and DM interaction terms.
   }
  \label{figure10}
\end{figure}

\begin{table*}[!htb]
\centering
\begin{tabular}{cc|c|ccc|ccc|l}
  \hline  \hline
  & & \hspace{4mm}$a$\hspace{4mm} &\hspace{4mm}$t_x$\hspace{4mm} & \hspace{4mm}$t_y$\hspace{4mm} &\hspace{4mm}$t_z$\hspace{4mm} &\hspace{4mm} $q_{xy}$\hspace{4mm} & 
  \hspace{4mm}$q_{xx}$ \hspace{4mm}& \hspace{4mm}$q_{yy}$\hspace{4mm} & \hspace{4mm} remarks \hspace{10mm}\\
  \hline
  &I     &  * & - & - & - & - & $\checkmark$ & $\checkmark$ &  \\
  &II    &  * & $\checkmark$ & $\checkmark$ & - &  * &  * & * & $|t_x|=|t_y|$  in a layer\\
  ${\bf B} \parallel [001]$ &III   & * & $\checkmark$/- & -/$\checkmark$ & - & - & * & * & Either $t_x$ or $t_y$ is nonzero. \\
  &IV   & $\circ$ & - & - & - & - & $\circ$ & $\circ$ & $q_{xx}=q_{yy}$ \\
  & V   & * & - & - & $\checkmark$/$\circ$ & - & * & * & $t_z$ depends on the layer stacking; $q_{xx}=q_{yy}$\\
  \hline
  &Z     & $\checkmark$ & - & * & - & - & $\checkmark$ & $\checkmark$ &$q_{xx} \simeq -q_{yy}$ in a layer  \cite{comm1} \\
  &Y     & - & $\checkmark$ & * & - & - & - & - & \\
  ${\bf B} \parallel [100]$&M   & $\circ$ & - & * & - & - & $\circ$ & $\circ$ &  \\
  &T   & $\circ$ & * & * & $\circ$ & $\circ$ & $\circ$ & $\circ$ & $|q_{xy}| \ll |a|, |t_z|, |q_{xx}|$, and $|q_{yy}|$ \\
  &S   & $\checkmark$ & * & * & $\checkmark$ & $\checkmark$ & $\checkmark$ & $\checkmark$ &  $|q_{xy}| \ll |a|, |t_z|, |q_{xx}|$, and $|q_{yy}|$ in a layer\\
  \hline
  &Z'  & - & * & * & $\checkmark$ & - & $\checkmark$ & $\checkmark$ & $q_{xx} = - q_{yy}$ \\
  ${\bf B} \parallel [110]$ &Y'  & - & $\checkmark$ & $\checkmark$ & - & - & - & - & $t_x=-t_y$ (${\bf t}\parallel [1\bar{1}0]$) \\
  &M'  & $\circ$ & - & * & - & $\circ$ & $\circ$ & $\circ$ & $q_{xx}  = q_{yy}$ \\
  \hline  
  $B=0$ & ($\theta > \theta_c$)	& - & - & - & - & - & $\checkmark$ & $\checkmark$ &  $q_{xx} = -q_{yy}$  in a layer\\
  $B=0$ & ($\theta < \theta_c$)	& $\circ$ & - & - & - & - & $\circ$ & $\circ$ &  $q_{xx}=q_{yy}$ \\
  \hline \hline
\end{tabular}
\begin{tabular}{rll}
\hspace{5mm} 
\end{tabular}
\caption{
  Cluster multipole decomposition of the spin configurations in each square cupola into 
  the monopole $a$, toroidal moment ${\bf t}$, and quadrupole tensor $q_{\mu \nu}$.
  See Eqs.~\eqref{eq:a}-\eqref{eq:q}.
  The symbol $\circ$ means that the net value for the four cupolas in the unit cell is nonzero in the CMF solutions;
  the symbol $\checkmark$ means that the value for each layer is nonzero, but the net value vanishes because of the cancellation between the layers;
  the symbol * means that the value for each square cupola is nonzero, but that of each layer vanishes because of the cancellation;
  the symbol ``-''  means that the value for each square cupola is zero.
  \label{tab:2}
}
\end{table*}


Next, we study the ground-state phase diagram by changing only the DM angle $\theta$ for the parameter set for SrTCPO. 
Such a change may be possible by a deformation of square cupolas, e.g., by an external pressure and chemical substitutions.
Figure~\ref{figure9} shows the results as functions of $\theta$ and the magnetic field $B$. 
In addition to the phases appearing in the previous sections (I, II, III, Z, Y, Z', and Y'),
we find additional phases IV, V, M, T, S, and M' in the small $\theta$ region.

When $B=0$, the system exhibits a phase transition at the critical angle $\theta_c=12.5\pm 0.05^\circ$
between the spin configuration of monopole type for $\theta<\theta_c$ [Fig.~\ref{figure10}(a)] 
and of $q_{x^2-y^2}$ quadrupole type for $\theta>\theta_c$ [Fig.~\ref{figure10}(b)]~\cite{kimura2016,kato2017} (see also Appendix~\ref{app:sc}).
This transition occurs mainly because of the energy competition between the DM interaction and the $J_1$ exchange interaction, as shown in Fig.~\ref{figure10}(c); the former energy increases while the latter decreases for $\theta>\theta_c$. 
We note that the $J_2$ exchange interaction also contributes to the stabilization of the monopole-type spin configuration. 
The competition is also understood from the spin configurations shown in Figs.~\ref{figure10}(a) and \ref{figure10}(b). 
For $\theta<\theta_c$, $\langle {\bf S}_i \rangle \times \langle {\bf S}_j \rangle$ is almost parallel to ${\bf D}_{ij}$, which is preferable for the DM energy, 
while the neighboring spin pairs are almost perpendicular to each other, which is unfavorable for the $J_1$ energy. 
They are vice versa for $\theta>\theta_c$.

Figure~\ref{figure9}(a) shows the phase diagram for ${\bf B}\parallel [001]$.
When turning on the magnetic field, the monopole ($q_{x^2-y^2}$ quadrupole) type spin configuration continuously develops into that of the phase IV (I) for $\theta < \theta_c$ ($\theta > \theta_c$). 
While increasing $B$, the phase IV is extended to the larger $\theta$ region, and instead the phase I is narrowed.
With a further increase of $B$, the phase IV turns into the phase V, while the phase I turns into the phase II before saturation in the region of $\theta \lesssim 110^\circ$.
The typical spin configurations are shown in Appendix~\ref{app:sc} and Supplemental Material for Ref.~\cite{kato2017}.

Figures~\ref{figure9}(b) and \ref{figure9}(c) show the phase diagrams for ${\bf B}\parallel [100]$ and ${\bf B}\parallel [110]$, respectively.
Similar to the case of ${\bf B}\parallel [001]$, by introducing the magnetic field ${\bf B}\parallel [100]$ and ${\bf B}\parallel [110]$,
the monopole ($q_{x^2-y^2}$ quadrupole) type spin configuration appears in the phase M (Z) and phase M' (Z') for $\theta < \theta_c$ ($\theta > \theta_c$), respectively.
However, the phases M and M' shrink as $B$ increases, in contrast to the case of ${\bf B}\parallel [001]$.
For ${\bf B}\parallel [110]$, the phase M' directly turns into the phase Z', whereas for ${\bf B}\parallel[100]$, 
intermediate phases S and T are found before entering to the phase Z. 
The typical spin configurations for these additional phases are shown in Appendix~\ref{app:sc}. 
In the intermediate $\theta$ region, the phase Z (Z') turns into the phase Y (Y') before saturation.

We summarize in Table~\ref{tab1} the antiferromagnetic order parameter ${\bf m}_{\rm AF}$  
and the interlayer-staggered component of the electric polarization, ${\bf P}_{\rm AF}$,
for the additional phases IV, V, M, T, S, and M' found in the small $\theta$ region in Fig.~\ref{figure9}.
We also show the net electric polarization ${\bf P}$ [Eq.~\eqref{eq:P}] in the table.

For $\theta < \theta_c$ at $B=0$, ${\bf m}_{\rm AF}$ is zero.
Accordingly, ${\bf m}_{\rm AF}$ remains zero in the phases IV, M, and M', where the spin configurations are continuously deformed from that for $B=0$
(see also Appendix~\ref{app:sc}).
Although the spin configurations drastically change through the transitions from IV to V and from M to T, ${\bf m}_{\rm AF}$ remains zero in both phases V and T.
On the other hand, ${\bf m}_{\rm AF} \neq 0$ in the phase S because of the antiferromagnetic layer stacking in contrast to the ferromagnetic one in the phase T.
In the phase V, within the numerical accuracy, the interlayer spins are uncorrelated in spite of the finite $J''$:
the states with ferromagnetic and antiferromagnetic layer stackings are energetically degenerate in the phase V.
We note that the spin configuration in the phase V is of toroidal type in terms of the cluster multipole decomposition discussed in Sec.~\ref{sec:multipole}.

For the electric property, remarkably, the net polarization ${\bf P}\parallel {\bf B}$ becomes nonzero in the phases IV, M, and M'.
As we will discuss in Sec.~\ref{sec:multipole}, this behavior originates from the monopole type spin configuration.
In the phase T, $P^y$ (perpendicular component to ${\bf B}$) becomes nonzero in addition to $P^x$ 
because the spin configuration is regarded as a superposition of a toroidal type and a monopole type spin configuration with the uniform manner of the layer stacking.
Meanwhile, in the phase S, ${\bf P}_{\rm AF} \perp [001]$ becomes nonzero because of the antiferromagnetic layer stacking of a similar mixed type spin configuration.

\subsection{Cluster multipole decomposition}\label{sec:multipole}

The ME behaviors in different phases found in the previous sections can be understood in terms of multipoles. 
In the present system, the multipoles are defined in a cluster form for a square cupola. 
For the cluster multipole description, we define a $3\times 3$ tensor by using the spin configuration in a square cupola as 
\begin{equation}
\mathcal{M}_{ij} \equiv \sum_{\ell} \tilde{r}^i_\ell S^j_\ell,
\end{equation}
where $i,j$ takes $x$, $y$, or $z$; $\tilde{{\bf r}}_\ell$ is the relative coordinate of site $\ell$ from the center of the square cupola, and the sum is taken for the four sites in the square cupola. Then, the tensor $\mathcal{M}_{ij}$ can be decomposed into the cluster multipoles, i.e., the pseudoscalar monopole $a$, the toroidal moment vector ${\bf t}=(t_x,t_y,t_z)$, and the quadrupole tensor $q_{ij}$, which are defined as
\begin{eqnarray}
  a   &=& \frac{1}{3}\sum_i \mathcal{M}_{ii},
  \label{eq:a}\\
  t_k &=& \frac{1}{2} \sum_{i,j} \varepsilon_{ijk} \mathcal{M}_{ij},
  \label{eq:t}\\
  q_{ij} &=& \frac{1}{2} \left( \mathcal{M}_{ij} + \mathcal{M}_{ji} - \frac{2}{3}\delta_{ij} \sum_k \mathcal{M}_{kk} \right),
  \label{eq:q}  
\end{eqnarray}
respectively~\cite{spaldin2008},
where $\delta_{ij}$ and $\varepsilon_{ijk}$ represent the Kronecker delta and the three-dimensional Levi-Civita symbol, respectively.

We summarize the results of the cluster multipole decomposition in Table~\ref{tab:2}. 
Here $q_{zz}$, $q_{yz}$, and $q_{zx}$ are omitted because $\tilde{r}^z_\ell = 0$ for all $\ell$ leads the three relations, $a=-q_{zz}$, $t_x=q_{yz}$, and $t_y = -q_{zx}$.
The nonzero components of the cluster multipoles explain the ME behaviors in each phase. 
For example,
in the phases I, Z, and Z', the nonzero ${\bf P}_{\rm AF}$ in ${\bf B}\parallel [100]$ and $[110]$ is naturally expected from the quadrupole of $x^2-y^2$ type, $q_{x^2-y^2} = q_{xx} - q_{yy}$.
The quadrupole also explains the divergent behavior of the dielectric anomaly in $\varepsilon^{[100]}(T)$ 
($\varepsilon^{[1\bar{1}0]}(T)$) for ${\bf B}\parallel [100]$ (${\bf B}\parallel [110]$) at the N\'eel temperature,
as commonly observed in BaTCPO~\cite{kimura2016,kato2017} [Figs.~\ref{figure4}(a) and \ref{figure4}(e)]. 
On the other hand, in the phase Y', the toroidal moment ${\bf t} \parallel [1\bar{1}0]$ becomes nonzero in each layer, 
which indicates the free energy has a coupling term between $E^{[001]}$ and $B^{[110]}$. 
This explains ${\bf P}_{\rm AF} \parallel [001]$ induced by ${\bf B} \parallel [110]$. 
Similarly, 
in the phase II, ${\bf t} \parallel [110]$ or ${\bf t} \parallel [1\bar{1}0]$ becomes nonzero in each layer, 
which explains ${\bf P}_{\rm AF} \parallel [1\bar{1}0]$ or $[110]$ induced by ${\bf B}\parallel [001]$. 
In the phase III, $t_x$ or $t_y$ becomes nonzero in each layer, which explains ${\bf P}_{\rm AF} \parallel [010]$ or $[100]$ induced by ${\bf B}\parallel [001]$. 

Meanwhile, in the newly-found phases in the small $\theta$ region, the net monopole $a$ is activated, together with the quadrupole tensor $q_{xx}=q_{yy} \neq 0$.  
This indicates that the free energy has a coupling term of $E^{\mu} B^{\mu}$ with a uniaxial anisotropy, i.e., 
the coefficient of $E^z B^z$ is different from that of $E^x B^x$ and $E^y B^y$.
This explains ${\bf P}\parallel {\bf B}$ in the phases IV, M, and M'. 
In the phase T, the net toroidal moment ${\bf t}\parallel [001]$ is activated,
which explains a nonzero component of ${\bf P}$ perpendicular to both ${\bf B}$ and ${\bf t}$, in addition to a component parallel to ${\bf B}$.  
In the phase V where ${\bf P}={\bf P}_{\rm AF}=0$, the nonzero ${\bf t}\parallel [001]$ indicates the free energy term of $E^x B^y - E^y B^x$. This means that ${\bf P}$ or ${\bf P}_{\rm AF}$, which is perpendicular to ${\bf B}$ and [001], is activated
by tilting the magnetic field from ${\bf B}\parallel [001]$ depending on the way of layer stacking.

\section{Summary and concluding remarks}\label{sec:summary}
In conclusion, we have investigated the magnetoelectric behavior of SrTCPO 
composed of antiferromagnetic square cupolas by the combination of experimental measurements and theoretical analyses.
In experiments, by the help of stable single crystal growth, we obtained the full magnetization curves 
at low temperature (1.4~K up to 45~T) for three different field directions, ${\bf B}\parallel [001]$, ${\bf B}\parallel [100]$, and ${\bf B}\parallel [110]$, and the dielectric constant as functions of temperature and the magnetic field (up to 18~T) for ${\bf B}\parallel [100]$ and ${\bf B}\parallel [110]$.
The magnetization curves show magnetization jumps, whose critical fields depend on the field direction,
similar to those of isostructurals BaTCPO~\cite{kimura2016a,kato2017} and PbTCPO~\cite{kimura2018b}.
The dielectric constant shows an anomaly at the critical fields. 
We found several differences between SrTCPO and previously studied BaTCPO and PbTCPO; in particular, 
the ratio of the critical field to the saturation field is much larger in SrTCPO for ${\bf B}\parallel [001]$.
To understand the experimental observations,
we studied a spin model by using the the CMF method, following the previous studies for BaTCPO~\cite{kato2017} and PbTCPO~\cite{kimura2018b}. 
We found that the model well explains all the data for SrTCPO, including the finite-$T$ phase diagrams, by tuning the model parameters.
The agreements strongly support the validity of the simple microscopic model and our analyses for the isostructural series of $A$TCPO.

We have also investigated further interesting ME behaviors by extending the model parameter space. 
Although we did not find any additional phases by linearly interpolating the parameters between the Sr and Ba cases, we unveiled a variety of unprecedented phases, including ferroelectric ones, by changing the DM angle with the parameter set for the Sr case.
We investigated the ME behaviors in all the phases, and rationalized them by using the cluster multipole decomposition. 
We found that the spin configurations in the additional phases for a small DM angle acquire the cluster form of not only quadrupole, which was already identified for the previous studies, but also monopole and toroidal moments. 
Thus, our results indicate that the antiferromagnetic square cupola could host all the multipoles giving rise to the linear ME effect.
A smaller $\theta$ is expected to be possibly realized by compressing the cupola in the [001] direction, e.g., by an external pressure and chemical substitutions. 
Our findings would stimulate further material investigation in the family of $A$TCPO
and the materials composed of the Cu-based square cupolas~\cite{hwu2002,giester2007,williams2015} 
for such intriguing ME behaviors.


\begin{acknowledgments}
The authors thank M.~Toyoda and K.~Yamauchi for fruitful discussions.
This work was supported by JSPS Grant Numbers JP17H01143, JP16K05413, JP16K05449
and by the MEXT Leading Initiative for Excellent Young Researchers (LEADER).
The ESR work was carried out at the Center for Advanced High Magnetic Field Science in Osaka University under the Visiting Researcher's Program of the Institute for Solid State Physics, the University of Tokyo.
Measurements of dielectric properties in a magnetic field were performed at the High Field Laboratory for Superconducting Materials, Institute for Materials Research, Tohoku University (Project No.~18H0014).
Numerical calculations were conducted on the supercomputer system in ISSP, The University of Tokyo.
K.K., M.A., M.H., S.K., T.K., and Y.M. are partially supported by JSPS Core-to-Core Program, A. Advanced Research Networks.
\end{acknowledgments}

\appendix
\section{Spin configurations in the small $\theta$ region}\label{app:sc}
\begin{figure*}[!htb]
  \centering
  \includegraphics[width=\textwidth]{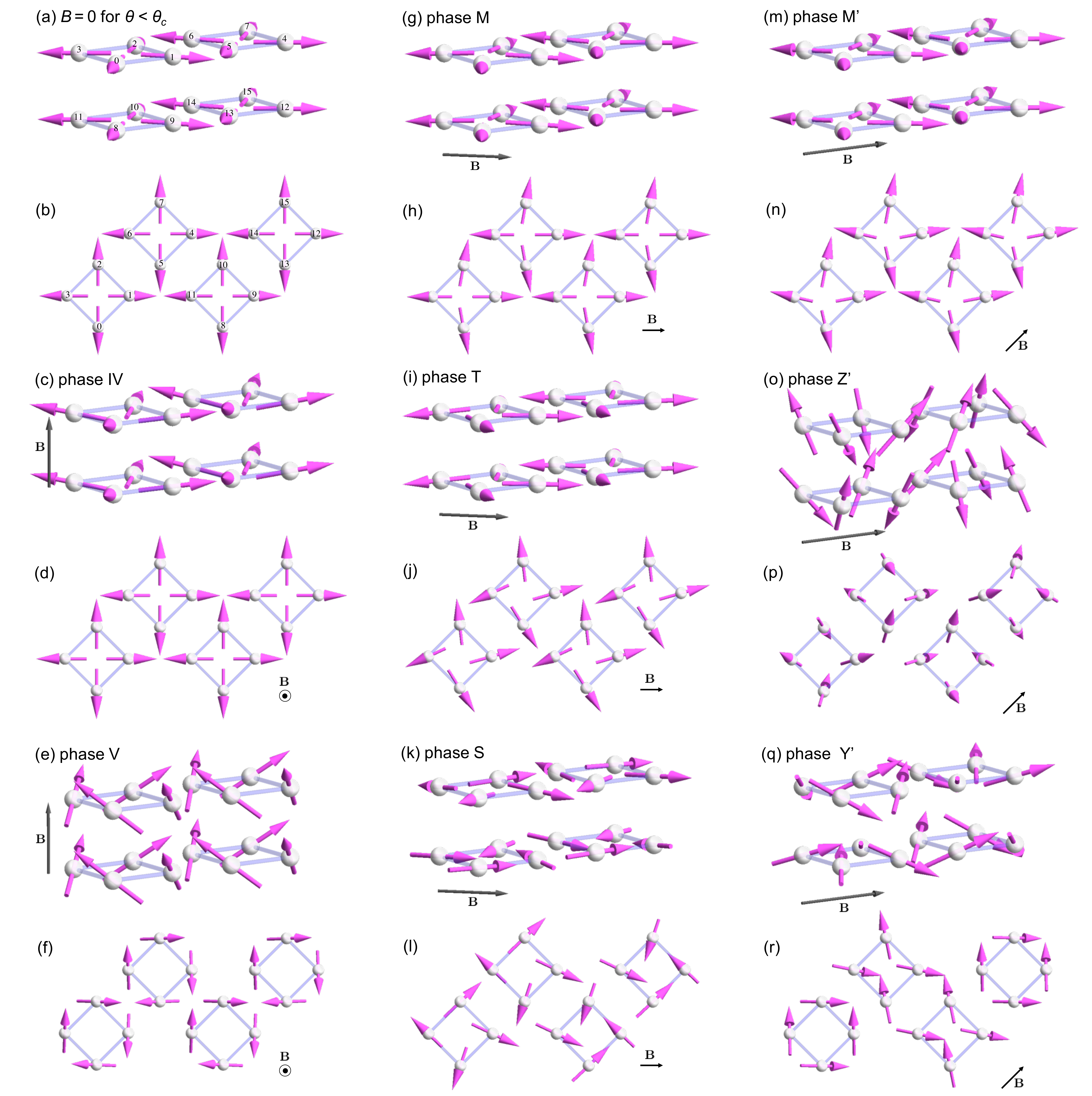}
  \caption{
  Spin configurations in the phases 
  (a,b) 
  $B=0$ for $\theta<\theta_c$,
  (c,d) IV,
  (e,f) V,
  (g,h) M,
  (i,j) T,
  (k,l) S,
  (m,n) M'
  (o,p) Z',
  and (q,r) Y'.
  (a,c,e,g,i,k,m,o,q) are the three-dimensional views, and
  (b,d,f,h,j,l,n,p,r) are the top views.
  The numbering in (a,b) 
  corresponds to Fig.~\ref{fig1}.
  They are obtained by the CMF method with
  the magnetic field (c-f) ${\bf B}\parallel [001]$, (g-l) ${\bf B}\parallel [100]$, (m-r) ${\bf B}\parallel [110]$ and
  the parameter set for SrTCPO (see caption of Fig.~\ref{fig2}) except for the DM angle $\theta$; 
  (a-n) $\theta=5^\circ$ and (o-r) $\theta=90^\circ$.
  The field strength is
  (a,b) $B = 0$,
  (c,d) $B = 0.5$,
  (e,f) $B = 1.5$,
  (g,h) $B = 0.3$,
  (i,j) $B = 0.4$,
  (k,l) $B = 0.5$,
  (m,n) $B=0.5$
  (o,p) $B = 0.8$,
and
  (q,r) $B = 1.5$,
   }
  \label{figA1}
\end{figure*}

Figure~\ref{figA1} shows typical spin configurations in the phases IV, V, M, T, S, M', Y', and Z'  in the small $\theta$ region, obtained by the CMF method.
Spin configurations of other phases 
(I, II, III, Y, and Z) have been reported in Ref.~\cite{kato2017}.

\section{Phase diagram for BaTCPO with ${\bf B}\parallel [110]$}\label{app:PD:BaTCPO}
\begin{figure}[!htb]
  \centering
  \includegraphics[width=\columnwidth,trim = 0 0 0 0,clip]{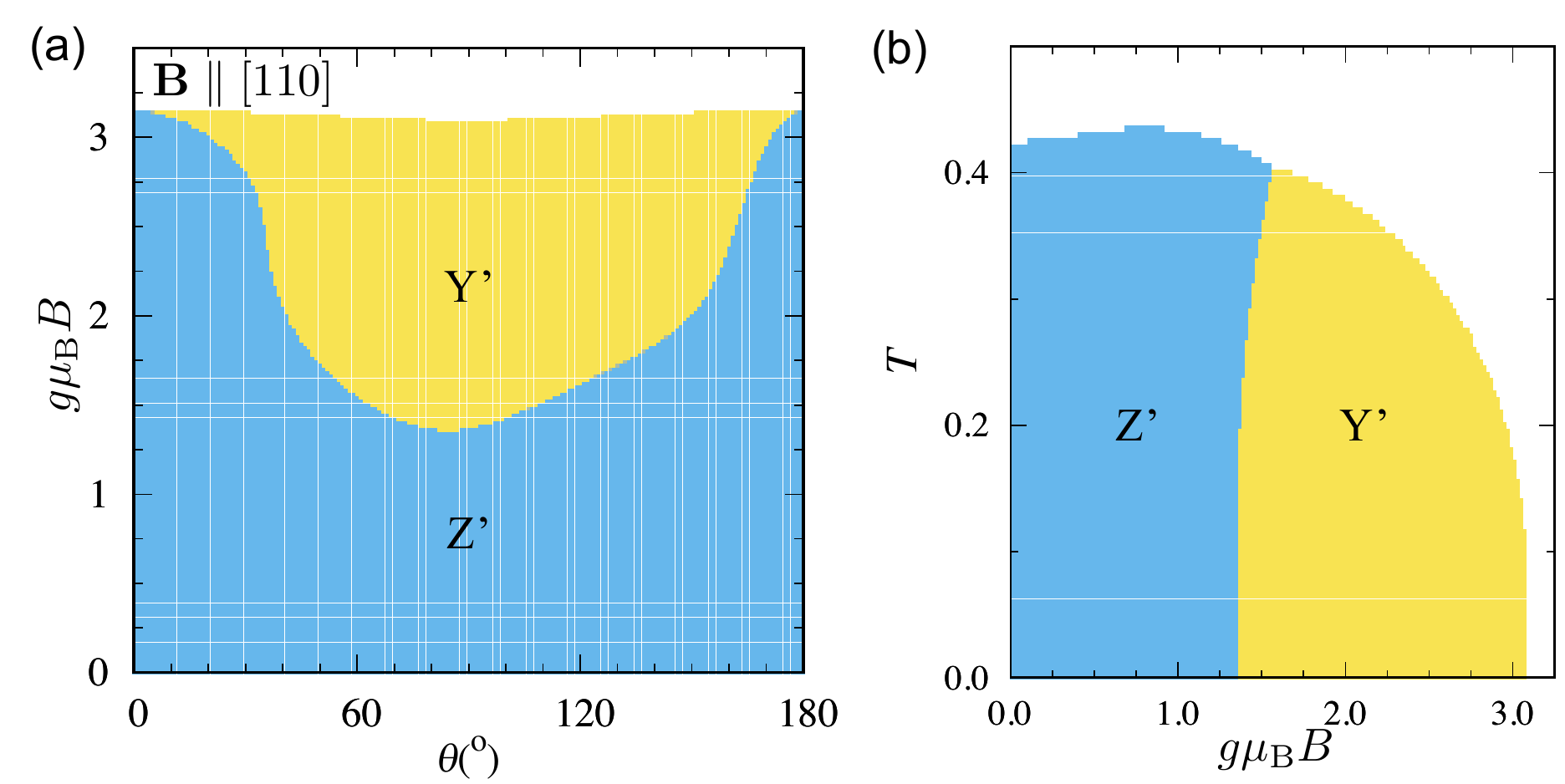}
  \caption{
  (a) Ground-state and
  (b) finite-$T$
  phase diagrams in the magnetic field parallel to $[110]$ (${\bf B}\parallel [110]$), 
  obtained by the CMF method with the parameter set for BaTCPO [Eq.~\eqref{eq:paramBa}]. 
  The DM angle is changed in (a), as in Fig.~\ref{figure9} for SrTCPO.
   }
  \label{figB1}
\end{figure}

Figure~\ref{figB1} shows the phase diagrams computed with the parameter set for BaTCPO [Eq.~\eqref{eq:paramBa}] 
and ${\bf B}\parallel [110]$ by the CMF method for comparison to 
those for SrTCPO in Figs.~\ref{figure9}(c) and \ref{figure6}(c).
The phase diagrams for the other two field directions were reported in Ref.~\cite{kato2017}.

\bibliographystyle{apsrev4-1}
\bibliography{cu4}

\end{document}